# A Single Sensor Based Multispectral Imaging Camera using a Narrow Spectral Band Colour Mosaic Integrated on the Monochrome CMOS Image Sensor

Xin He,[1] Yajing Liu,[1] Kumar Ganesan,[2] Arman Ahnood,[3] Paul Beckett,[3] Fatima Eftekhari,[4] Dan Smith,[4] MD Hemayet Uddin,[4] Efstratios Skafidas,[1] Ampalavanapillai Nirmalathas,[1] and Ranjith Rajasekharan Unnithan[1]

[1]Department of Electrical and Electronic Engineering, The University of Melbourne, Melbourne, VIC, 3010, Australia. Correspondence and requests for materials should be addressed to R.R.U.
(email: r.ranjith@unimelb.edu.au)
[2]School of Physics, University of Melbourne, Melbourne, VIC, 3010 Australia
[3]School of Engineering, RMIT University, Melbourne, VIC, 3000, Australia.
[4]Melbourne Centre for Nanofabrication, Australian National Fabrication Facility, Clayton, VIC, 3168, Australia.

Abstract

A multispectral image camera captures image data within specific wavelength ranges in narrow wavelength bands across the electromagnetic spectrum. Images from a multispectral camera can extract additional information that the human eye or a normal camera fails to capture and thus may have important applications in precision agriculture, forestry, medicine and object identification. Conventional multispectral cameras are made up of multiple image sensors each fitted with a narrow passband wavelength filter and optics, which makes them heavy, bulky, power hungry and very expensive. The multiple optics also create image co-registration problem. Here, we demonstrate a single sensor based three band multispectral camera using a narrow spectral band RGB colour mosaic in a Bayer pattern integrated on a monochrome CMOS sensor. The narrow band colour mosaic is made of a hybrid combination of plasmonic colour filters and heterostructured dielectric multilayer. The demonstrated camera technology has reduced cost, weight, size and power by almost n times (where n is the number of bands) compared to a conventional multispectral camera.

Introduction

In a conventional CMOS based image sensor, colour imaging relies on the integration of filters on top of the photodetector array [1-5]. These filters typically cover the three primary colours (bands), red, green and blue (RGB), predominately in a Bayer pattern [6-9]. As the human eye is more sensitive to green light than either red or blue, the widely used Bayer filter mosaic is formed from alternating rows of red-green and green-blue filters with twice as many green as red or blue filters. Three different materials are used for producing the primary colours with wide spectral bands (spectral width of around 90-100 nm) for all wavelengths [6, 9]. Multispectral cameras extend this concept to capture images with multiple colour bands and with narrow pass bands (i.e., narrow spectral widths) [10-12]. Images from a multispectral camera can extract significant amount of additional information that the human eye or a normal camera fails to capture and thus have important applications in precision agriculture, forestry, medicine, object identifications and classifications [10-15]. Conventional multispectral cameras are made up of multiple CMOS sensors each externally fitted with a narrow passband wavelength filter. For example, three bands would require three image sensors with associated electronics, three narrow bandpass filters and three optics. Depending on the application, the spectral width measured at the FWHM (Full Width Half Maximum) of a multispectral imaging camera varies between 10 nm and 90 nm [10-12]. Figure 1S and Table 1S (supplementary information) show comparison between a conventional multispectral camera with six bands and a single sensor based multispectral camera. The need for multiple sensors for each band results in a number of problems. Firstly, it means that multispectral cameras tend to be bulky and power hungry, which in turn limits their wider deployment in portable applications such as drone-based precision agriculture or for hand- held and portable uses such as wound monitoring, vein detection and skin screening [13-15]. It also greatly complicates the task of optical alignment which ensures that precisely the same scene is captured across all bands. Further, image co-registration problems will emerge from a slight mismatch between images in each band, which will require complex image processing to correct [16-17]. The above problems can be eliminated if a multispectral camera is developed using one single image sensor. This has prompted research in developing single sensor based multispectral cameras [18-19].

Conventional pigments and dye-based filters are not suitable for making narrow band filters mosaic because their spectral widths tend to be too large (90-180nm) and hence preventing the use of a normal colour camera for multispectral imaging applications [11,20]. Furthermore, these pigments tend also to be sensitive to UV radiation, degrade at high temperatures and are not particularly environmentally friendly.

Conventional technology for making a single narrow bandpass filter requires deposition of several layers of different dielectric materials (for example, 40 layers [21]) with precise thickness. Here, for making each narrow colour band requires different thickness combinations for these 40 layers. For example, three narrow colour bands require three different filters with each colour band repeated thousands of time to form a filters mosaic on a CMOS sensor using the conventional technology for getting an image. This will require thousands of steps for laying down thin films of precise thickness for each band separately with multiple complex masking and alignment processes with a large failure rate that significantly increases manufacturing costs and complexity [20]. As a result, it is extremely difficult to develop a low cost colour mosaic with narrow pass bands integrated onto the pixels of a CMOS chip.

Advancement in nanofabrication has enabled fabrication of novel nanophotonic devices including colour filters [6, 22-26]. Plasmonic colour filters with different geometries have proven their ability to tune the wavelength from UV to NIR range [26-48]. Furthermore, plasmonic filters are shown to be suitable for developing one single narrow band filter in the visible wavelengths [46-47] and short wavelength infrared (SWIR) [48]. In these filters, spectral FWHM values of 30nm (visible) and 80nm (SWIR) were achieved in selected wavelengths. 1D metal grating filters have been explored with an aim to reduce the spectral width of plasmonic colour filters. Here, the plasmons are excited through an interface between the metal and dielectric [32, 46-48]. A spectral width of 64 nm was demonstrated using a 1D metal grating filter made of silver nanoslit array fabricated on a glass substrate [48]. In addition, a 1D metal grating filter based on Al nanoslit integrated onto a thick Al2O3 buffer layer has been demonstrated with spectral widths of 20nm [32]. However, it has been proved both theoretically and experimentally [33, 49] that 1D metal grating filters are polarization dependent and can produce colours only under transverse magnetic (TM) illumination. 2D metal grating filters and dielectric guided mode resonance (GMR) filters are reported to bypass the polarization effects [25, 50-52]. A hexagonal hole array inserted in a metal-dielectric-metal multilayer has been reported to slightly reduce the spectral width [20]. Overall, it appears that most of the narrow-band spectral filters reported are unsuited for use in multispectral image sensors for a range of reasons including: inability to achieve narrow spectral width over a wide range of wavelengths, fabrication complexity to achieve narrow multi-band filters mosaic, polarization sensitivity and the requirement for TM illumination in the case of 1D metal grating filters.

Here, we demonstrate a single sensor-based multispectral camera using a hybrid narrow spectral band RGB colour mosaic fabricated on a quartz substrate and then integrated on a monochrome CMOS image sensor. The presented filter mosaic can be easily tuned to any wavelength and requires only one processing stage to derive multiple bands. The filter mosaic consists of a hybrid combination of a double sandwich of silicon nitride-silica-silicon nitride layers (heterostructured dielectric multilayer) covered by a hole array pattered in CMOS compatible aluminium (the plasmonic filter). The heterostructured dielectric multilayer is a common base layer for all the bands and the thickness values were optimized to reduce the spectral width in a given wavelength range of interest. A single layer of plasmonic filter is used for wavelength tuning. The mosaic on quartz is then integrated onto a monochrome CMOS image chip (Sony) using a flip-chip bonder resulting in a 3 cm x 3 cm size, three-bands single-sensor based multispectral image camera. The performance of the camera is first demonstrated using a standard Macbeth Chart. It is then fitted onto a lightweight DJI Phantom 3 drone to demonstrate its imaging capabilities in the field and for making handheld sensors. Because only a single camera chip is required, weight, power and complexity can be reduced by a factor of n, where n is the number of bands compared to the conventional approach using multiple cameras.

Results
Design and optimization of the hybrid colour mosaic
The RGB hybrid filter mosaic was designed and optimized by 3D simulation within the finite element based COMSOL Multiphysics® package. Each filter geometry was investigated separately using a 10nm wavelength step size. Figure 1(a) shows 3D simulation model of the colour mosaic with 6 layers. The basic unit cell for this simulation encompassed the diamond shaped pattern of holes highlighted in red in the top of Figure 1(a). The simulation model, Figure 1 (a) consisted of a 150 nm thick layer of aluminium patterned with a hexagonal arrangement of holes (the plasmonic filter: 1 layer) over repeating layers of $Si_3N_4$ - $SiO_2$ – $Si_3N_4$ (heterostructured dielectric multilayer: 5 layers), deposited on a semi-infinite glass substrate (n=1.5). A 200nm layer of Spin-On-Glass (SOG) was assumed to cover the aluminium layer for index matching and to avoid any shorting with metallic pads on CMOS chip while integration. Finally, a perfectly matched layer (PML) was used at the top and bottom of the model to avoid the effects of reflected light in the transmittance spectrum and periodic boundary conditions were applied to the four sides to allow a large area to be simulated without costing excessive memory and time.

The pitch and hole diameters were varied to obtain peak transmission at 440 nm, 530 nm and 625 nm [29] using the plasmonic layer. The objective was then to determine and validate the wavelength at which maximum transmittance occurs for these filters. Light was excited from the aluminium side (top side) using port boundary conditions and S-parameters used to find the transmittance ($|S_{21}|^2$) of the filters. As in [53], refractive index values of 1.42 and 1.5 were used for the SOG layer and quartz substrate, respectively. Filmetrics was used to experimentally determine a refractive index for $SiO_2$ of 1.45 and around 1.9 for the $Si_3N_4$ and then used in the simulations. As shown in Figure 1 (b) – (d), the plasmonic layer has produced the required transmission peak (colour), but with large spectral width of 70 nm, 95 nm and 160 nm for blue, green and red respectively.

The plasmonic layer was then combined with a heterostructured dielectric multilayer with 5 layers made of double $Si_3N_4$ (230 nm) - $SiO_2$ (350 nm) - $Si_3N_4$ (230 nm) sandwich layers to form the hybrid filter with optimized thickness values to reduce the spectral width. The heterostructured dielectric multilayer with a constant thickness values form a common base layer for all RGB bands in the filter mosaic. In the hybrid filter, one single plasmonic layer has removed most of the spectral contents on either side of the peak transmission wavelength in a given range of wavelengths. This has significantly reduced the requirement of large number of layers in the multilayer to produce narrow bands. Figure 1 (b) - (d) shows the simulated spectra of the red, green and blue filters in the mosaic from 400nm to 900 nm along with the electric field distributions at peak wavelengths. Here, the thickness of the $Si_3N_4$ - $SiO_2$ - $Si_3N_4$ base layer and the plasmonic layer (Al) are kept constant and the pitch (period) and diameter of the holes in the 150 nm aluminium plasmonic layer varied to tune the red, green and blue filters. The optimized hybrid filter mosaic parameters are given in the supplementary information (Table 2S). The FWHM of the hybrid red filter was reduced to 35nm from this simulated spectrum. For the green and blue filters, the FWHM was reduced to 30nm and 17nm respectively. Furthermore, this topology has considerably reduced the fabrication complexity as the thickness of the layers can be kept constant when fabricating the narrow band mosaic, and wavelength tuning can achieved by varying the pitch of the holes in the top single nanoscale thick plasmonic layer. The resonance peak shift with respect to different angle of incidence (0 – 80 degrees field of veiw, FOV) was estimated for the hybrid filter (green hybrid filter was taken as an example) as shown in the supplementary information (Figure 2S). The resonance peak position remains almost constant irrespective of the angle of incidence with slight decrease in the transmission intensity.This FOV is in the acceptable limit with suitable optics attached to the multispectral camera.

## Materials and methods
### Fabrication of the hybrid colour mosaic

The hybrid filter mosaic was fabricated on a 4-inch quartz wafer. The fabrication process is shown in Figure 2. Firstly, the wafer was cleaned using Acetone and IPA (Isopropyl alcohol) with ultrasonic agitation followed by 2 minutes of plasma pre-cleaning. The wafer was then deposited with a-Si3N4 and a-SiO2 by Plasma enhanced CVD (Oxford Instruments PLASMALAB 100 PECVD). The circular 4-inch wafer was diced into 2 cm x 2 cm square pieces and the centre pieces were selected for further fabrication due to their uniformity of Si3N4 and SiO2 film thicknesses. The measured refractive index of Si3N4 and SiO2 developed by PECVD were 1.9 and 1.45, respectively. The deposition rates were optimized and approximately 23nm/min (10% tolerance) and 70nm/min with the composition shown in the supplementary information Table 3S.

The thickness of $Si_3N_4$ was optimized to be 230nm and the $SiO_2$ to 350nm. Starting with $Si_3N_4$, a total of five layers of $Si_3N_4$ and $SiO_2$ were deposited on the quartz wafer. After fabricating the multilayer structure, a 150 nm thick aluminium layer was deposited on the top of the structure using an E-beam evaporator (Intlvac Nanochrome II) at a constant rate of 0.2Å/Sec. An ellipsometer was used to measure the refractive index of the aluminium and the result showed it fits Rakić's experiment [54]. These data were subsequently used in the simulation model. A metallic nanohole array comprising varying pitch and hole diameters was fabricated on the aluminium film using EBL lithography process and deep reactive ion etching using the optimized values from simulations (Table 2S). A thin ZEP 520A resist was spin coated on the device at 1500rpm for 1.5 minutes, followed by 5 minutes baking at 180℃. The pattern was exposed by EBL (Vistec EBPG5000plusES) with 1.5nA current and 400um aperture for 4 hours. The sample was then developed in n-Amyl acetate for a minute followed by a rinse with IPA and DI water. The exposed pattern was etched by Deep reactive-ion etching (DRIE Oxford Instruments PLASMALAB100 ICP380) at 40℃ with forward power of 1000W and 20sccm $Cl_2$ under 2mT chamber pressure for 40 seconds to form the holes. ZEP resist was then removed by DRIE at 40℃ with a forward power of 1000W and 50sccm $O_2$. Finally, the Spin-On-Glass [53] was spin coated on the top surface at 4000rpm for 20 seconds, followed by baking on a hotplate at 210°C for 10 minutes.

## Discussion
### Spectrum measurement and discussion.

The fabricated hybrid colour filter on the quartz substrate was cut into 2×2 $mm^2$ squares using a dicing saw. The dicing step is carried out using G1A flange blade, and is optimized with hairline alignment. This alignment can adjust the cut on the substrate to the center of the hairline to precisely dice the edge with minimal edge damage. Hence, the dicing has not affected optical performances of the sensor. Figure 3 (a) shows optical images of the RGB filter mosaic in transmission mode under an optical microscope (Olympus BX53M) with 40× magnification. SEM image of a section of the hybrid filter mosaic from top view is shown in Figure 3 (b). One unit size of the hybrid RGBG mosaic is 11.2 µm×11.2 µm. The spectra of the hybrid RGB filters were measured using a CRAIC spectrometer (Apollo Raman™ Microspectrometer) and CytoViva Hyperspectral Imaging in transmission mode. Figure 3 (c) shows the RGB spectrum from 400nm to 900nm. The spectral sensitivity (responsivity) of most commercial image sensors working in the visible and near-IR have different sensitivity with respect to wavelengths (Figure 3S, supplementary information). Hence the experimentally measured spectra of the RBG was multiplied with responsivity versus wavelength for the image sensor to get the actual spectra as shown in Figure 3S (b). Figure 3 (f) shows the CIE chromaticity chart overlaid with the transmission data, demonstrating that the RGB filter values are falling in the appropriate part of the colour space. There is a small shift of green towards yellow due to a minor secondary peak in the green transmission spectrum. However, this small shift in the green coordinate is still falling around the achromatic point and is within acceptable limits. The experimental transmission efficiency for RGB filters in the mosaic are around 10 %. The low transmission efficiency is compensated by making each filter band of size 11.2 um covering 2 by 2 pixels (one pixel size: 5.6 um) to increase the light absorption by the photodetectors and hence to increases the signal content. Further compensation can be achieved by increasing exposure time of the image sensor in low light conditions while capturing the images. The FWHM of the red filter in the mosaic has the best performance with a width of around 45nm, while the blue and green filters exhibit FWHM values of 60nm and 60nm, respectively. The measured FWHM values are slightly wider than the results obtained from computer simulations for two primary reasons. Firstly, variations in the deposition rate of $Si_3N_4$ and the fact that $SiO_2$ growth using PECVD has larger tolerances than in the E-beam evaporator. While a high temperature (250°C) during the deposition of $Si_3N_4$ and $SiO_2$ results in a good quality of dielectric, it restricts the available methods for verifying the exact deposition thickness to the Filmetrics software sensor system in the PECVD, which is less accurate than AFM. Secondly, as shown in the SEM image of Figure 3 (b), the pitch and the hole shape in the plasmonic layer can vary due to fabrication tolerances (such as minor under cut in holes and nanoscale thickness variations) from the ideal (simulated) case. The cross talk among pixels is reduced by mounting the filter mosaic upside down to minimise the effect of substrate thickness. This prevents the off normal incident light of one pixel entering the neighbouring pixels. Further cross talk reduction was achieved by making each filter band of size 11.2 um covering 2 by 2 pixels (one pixel size: 5.6 um) to increase the light absorption by the photodetectors and hence to increases the signal content.

Integration of the hybrid colour mosaic onto a CMOS image sensor
The narrow band filter mosaic was then integrated onto a CMOS chip using a flip chip bonder (Figure 4S) for accurate alignment as shown in Figure 3 (d). The top of the filter mosaic was coated with SOG to match the refractive index (thus increasing the transmission) and also to reduce the spectral width as well as preventing shorting the sensor while integration (the hybrid filter was integrated on the image sensor upside down to avoid cross talk). The image sensor used was SONY ICX618 with pixel size of 5.6μm and resolution of 0.3 Megapixels (640×480). The image sensor protective glass was removed for the filter integration (Figure 5S). To compensate for the low transmission of the filters and also to increase the light absorption in photodetectors (pixels), each filter in the mosaic covered a 2×2 block of photodetectors in the CMOS image sensor, resulting in 160×120 pixels per band. For the integration, the PMMA based homemade adhesive (the PMMA powder was diluted in a small amount of anisole, followed by staying in the 100 class cleaning room for two weeks for making the adhesive) is used an between the filter mosaic and the image sensor while doing the integration using flip chip bonder. Here, we firstly spin coat a thin layer of PMMA on top of the image sensor and then integrate the filter on it after aligning with flip-chip bonder (supplementary Figure 4S). The performance of the colour filter was also verified by integrating on another image sensor MT9P031 (Figure 5S).

The optics used for the camera has *f* number 1.4 with *f* = 6mm (*f*-focal length) and the developed single sensor based multispectral camera is shown in Figure 3 (e). The camera was characterized using a 24-patch Macbeth Color Checker as an object (Figure 4 (g)). 8bit multispectral raw object data was captured by the camera and then transmitted to a laptop for image processing using MATLAB as shown in Figure 4 (a). Figure 4 (f) shows a plot of the signals from the pixels across the transect indicated by the dashed red line across the Macbeth Chart in Figure 4 (a). The dotted red line spans across the grey colour patches on the Macbeth colour Checker and which shows the pixel intensity variations are captured in the raw image. A *demosaicing* algorithm was used to extract red, blue and green channels from the multispectral raw data of the Macbeth Chart (Figure 4 (b)). The red, green and blue channels were recombined to get a colour image as shown in Figure 4 (c). Due to initial uncertainty of the RGB colour balance, colour correction and white balancing were required. Figure 4 (d) and (e) show images after colour correction and white balancing, respectively.

Figure 4 demonstrated that each band can be retrieved from the 8 bit multispectral raw image data. Another requirement of the multispectral image camera is extraction of each narrow band and then overlay for different spectral band combinations (red-green (RG), Green-Blue (GB) etc). Figure 5 shows that the recovered Macbeth Chart from Figure 4 (e ) can be used to recover different multispectral band combinations such as RG, RB and GB (Fig 5 (b)) which is desirable in many applications for finding NDVI (normalised differential vegetation index) for precision agriculture to find plant diseases [9-10], finding required information in a band for object identification and also in finding emissions in a narrow band for biomedical applications. Figure 5 (c) shows the CIE chart of the recovered Macbeth Chart. The chart demonstrates that the recovered colour values are falling in the appropriate part of the colour space in comparison to a standard CIE chart of the Macbeth Chart (Figure 6S, supplementary information).

The single sensor based multispectral camera was mounted on a DJI Phantom 3 for testing the sensor performance from a real aerial platform in an outdoor environment (an outer urban park) as shown in Figure 6 (a) and 6 (b). The sensor was mounted without gimbal as shown in Figure 6 (a). Figure 6 (c) shows the raw multispectral images of the Macbeth colour chart on the ground captured from a 15 m height above the ground, the other patches are calibration images. Figure 6 (d) shows pixel intensity values across a line over the white and black crossing and the Macbeth Chart images captured by the single sensor based multispectral camera. The pixel intensity variations with respect to positions are consistent with colour intensity variations in the Macbeth Chart and black and colour variations without any white balance. From the intensity variations, it was demonstrated that the image clarity was well within acceptable limits. Furthermore, the suitability of the raw multispectral image for making handheld sensors for precision agriculture was demonstrated by creating a Red-Blue vegetation index (RBVI), (R-B)×255/(R+B). An area with green grass and dry grass was captured by holding the drone mounted camera 1.5 m above the ground and the recorded multispectral raw image is shown in Figure 6 (e). R, G and B individual bands were recovered from the raw multispectral image to get RBVI as shown in Figure 6 (f). RBVI image shows that area of high dense green grass (red colour in the image) compared to dry grass (blue colour) which demonstrates the capability of the sensor platform in real applications.

In conclusion, the paper demonstrated a single sensor based narrow band multispectral imaging using a hybrid RGB colour mosaic integrated onto a CMOS sensor. The colour mosaic was designed in such a way that multiple bands can be fabricated on a quartz wafer in a single run and offers easy tuning of colours, in contrast to conventional techniques that demand several independent runs with complex alignment processes. The hybrid filter mosaic was made of a heterostructured dielectric multilayer structure consisting of a $Si_3N_4$ – $SiO_2$ - $Si_3N_4$ sandwich as a common base layer for the filter mosaic to reduce the spectral width followed by a metal layer made of aluminum film perforated with holes on the base structure as a plasmonic layer. The colour tuning is achieved by varying the pitch of holes and hence can be fabricated in a single run with no complex alignment required for the different bands. The thickness values required for the base and the plasmonic layers were optimized to obtain narrow spectral widths. The spectral widths of the RGB mosaic are 60 nm, 60 nm and 45 nm for the red, green and blue, respectively. The mosaic is then integrated onto a Sony sensor using a flip-chip bonder for better alignment accuracy with a thin layer of PMMA for adhesion, and refractive index matching. The single sensor based narrow multispectral imaging capability was demonstrated using a Macbeth colour chart followed by retrieving individual bands using demosaicing techniques and their combination to retrieve the Macbeth Chart after colour correction and white balance. The sensor was then fitted onto a lightweight DJI Phantom 3 Drone to demonstrate its imaging capabilities in a field using RB Vegetation Index. Because only a single sensor chip was used for the camera, it required only around one third of the weight and power of a conventional multispectral camera. In general, weight, power and complexity were reduced by a factor of n times, (where n is the number of bands) compared to a conventional multispectral camera using multiple sensors, electronics and optics. The demonstrated sensor will have applications in drone-based imaging for precision agriculture, developing portable low-cost sensors for wound healing, blood vein detection, mining and forensic applications.

Acknowledgements
This work was performed in part at the Melbourne Centre for Nanofabrication (MCN) and at the RMIT Micro Nano Research Facility (MNRF) in the Victorian Node of the Australian National Fabrication Facility (ANFF). We acknowledge the help from Mr. Bryce Widdicombe for drone tests. The authors acknowledge financial support from the Australia Research Council under Discovery Project: DP170100363. This work was performed in part at the Materials Characterisation and Fabrication Platform (MCFP) at the University of Melbourne and the Victorian Node of the Australian National Fabrication Facility (ANFF).

Conflict of interests
The authors declare that they have no conflict of interest.

List of Figures

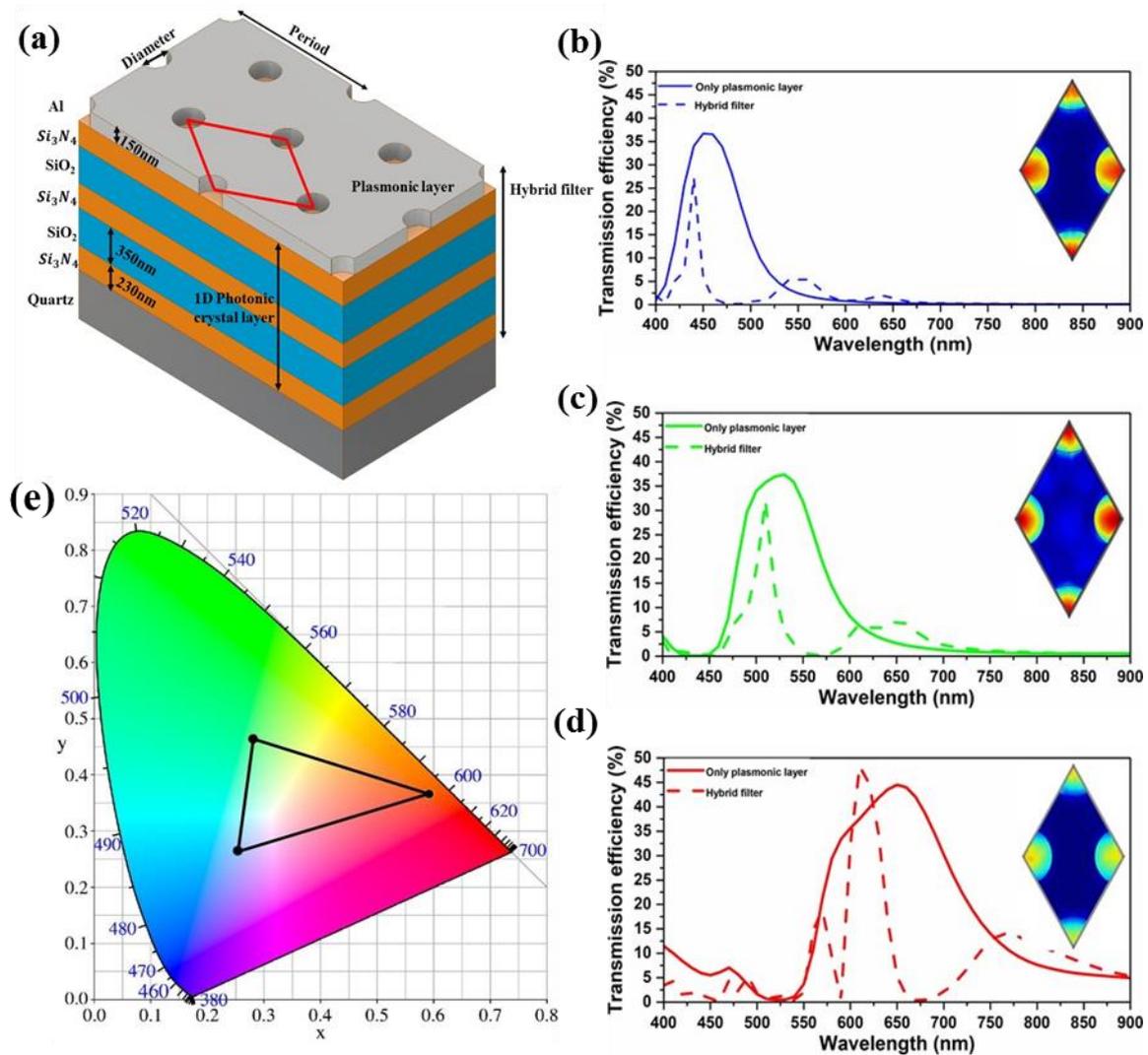

Figure 1: (a) 3D simulation model of hybrid filter in the colour mosaic. The hybrid filters consist of 6 layers made of double $Si_3N_4$ (230 nm) - $SiO_2$ (350 nm) - $Si_3N_4$ (230 nm) sandwich layers that forms a common base multilayer structure (heterostructured dielectric multilayer: 5 layers) and a 150 nm thick aluminium perforated with hexagonal arrangement of holes (plasmonic layer:1 layer). (b) Numerically simulated transmission spectra of the blue plasmonic layer and blue hybrid filter showing the reduction in spectral width. The wavelength was swept from 400 nm to 900 nm. The spectral width at full width at half maximum (FWHM) of the hybrid blue filter was reduced to 17 nm from 70 nm produced by the plasmonic aluminum layer. Subset image shows normalized electric field at the peak wavelength of 440 nm (c) The spectral width (FWHM) of hybrid green filter was reduced to 30 nm from 95 nm produced by the plasmonic layer. Subset shows normalized electric field at the peak wavelength of 530 nm (d) The spectral width (FWHM) of the hybrid red filter was reduced to 35 nm from 160 nm produced by the plasmonic layer. Subset shows normalized electric field at the peak wavelength of 625 nm (e) CIE chromaticity chart of simulated blue, green and red hybrid colour filters in the mosaic.

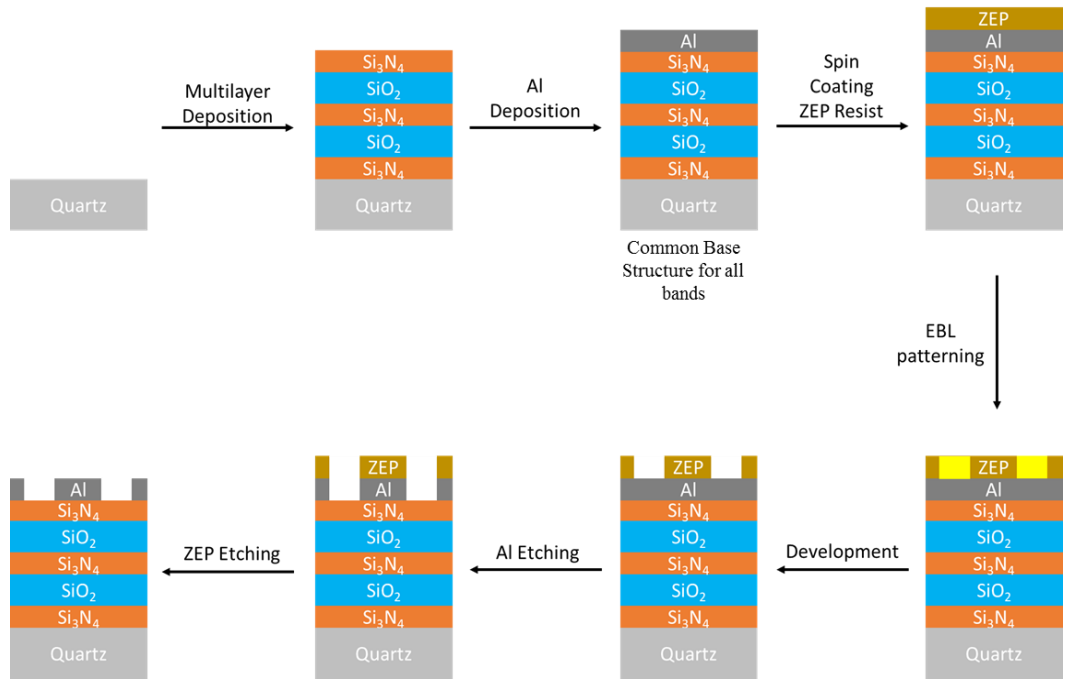

Figure 2. Fabrication process for our proposed narrow band filters mosaic with common base.

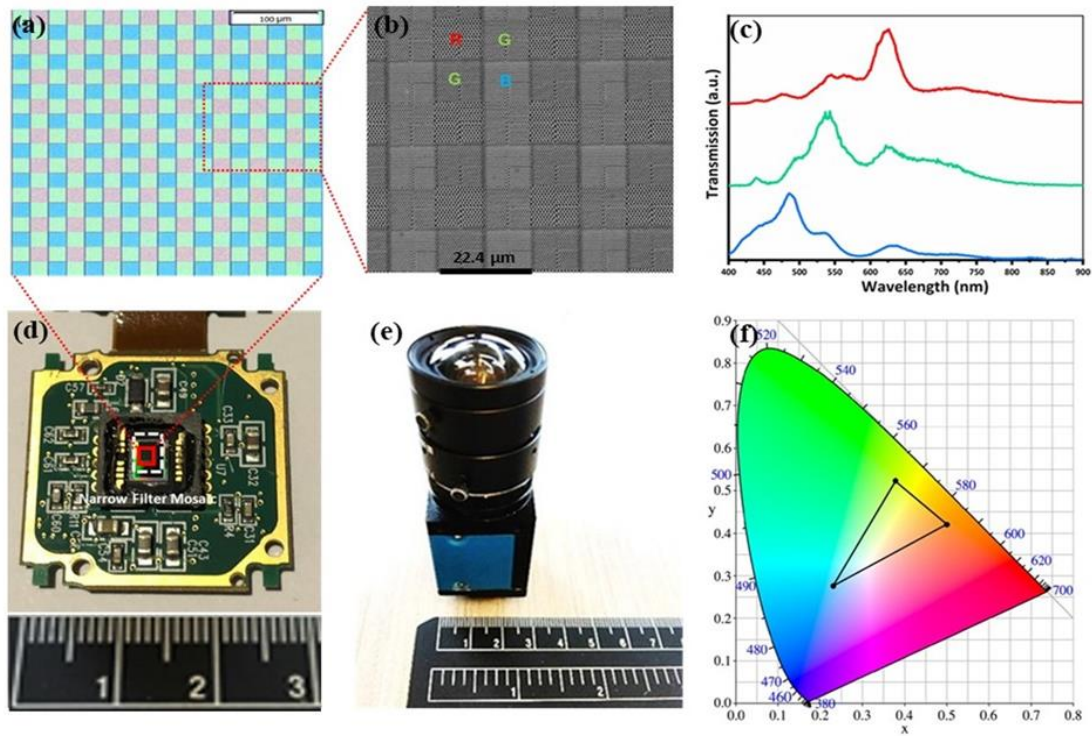

Figure 3: Integration of the hybrid narrow spectral band mosaic on the image sensor (a) Image of the colour mosaic under optical microscope in transmission mode with magnification × 40 (b) SEM image of a section of the hybrid mosaic from top view. One narrow band RGBG unit size is 11.2 µm×11.2 µm (c) Experimental transmission spectra of the narrow spectral band red, blue and green (RGB) colour filters from the colour mosaic. The spectral widths (FWHM) of RGB are 45 nm, 60nm and 60 nm, respectively (d) The colour mosaic integrated on SONY ICX618 sensor pixels using a flip chip bonder for alignment (size 3 cm× 3 cm) (e) The single sensor based multispectral camera. The mosaic integrated image sensor fitted with housing and optics with $f$ number 1.4 for multispectral imaging (f) CIE chromaticity chart of the hybrid narrow band blue, green and red colour filters in the colour mosaic from experimental spectra.

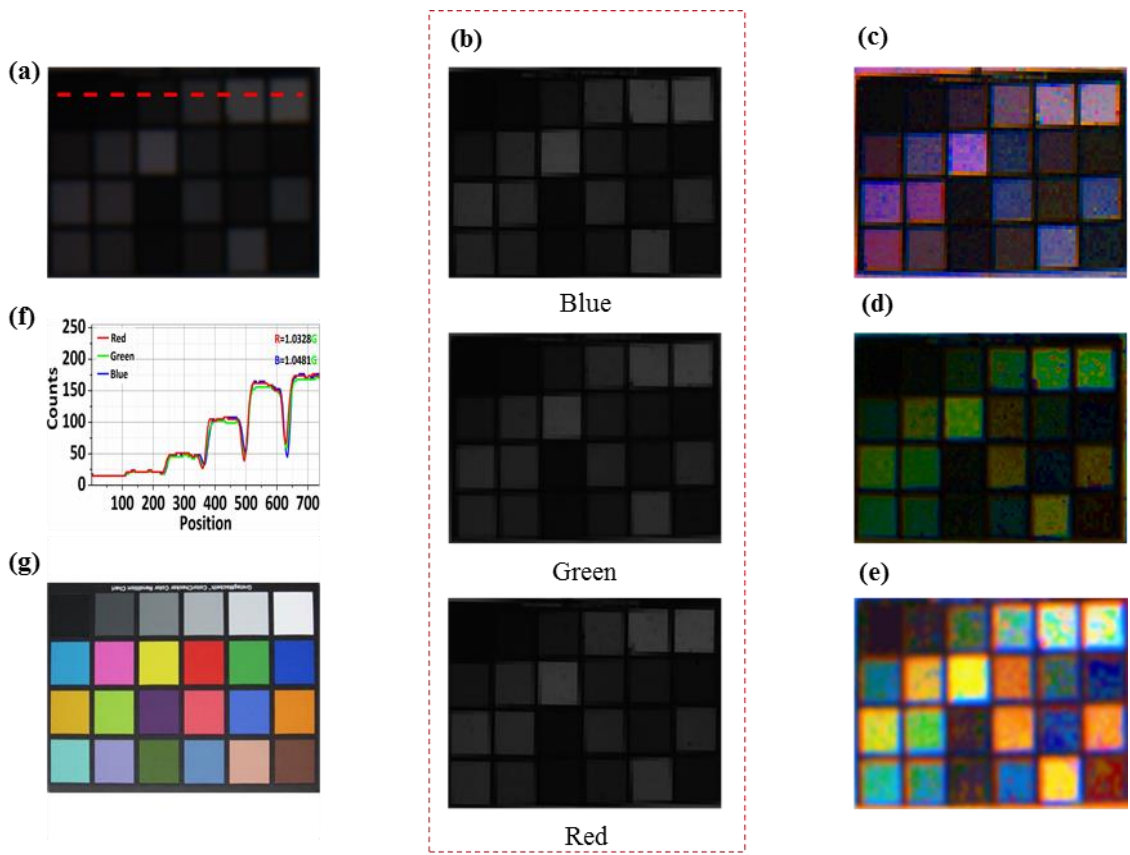

Figure 4: Image reconstruction process from the single sensor based multispectral camera (a) 8 bit multispectral raw data of 24 - patch Macbeth Chart captured by the multispectral camera (b) Three narrow wavelength channels (RGB) extracted from the raw image (c) The three channels are recombined to get a RGB colour image (d) The colour image after colour correction and (e) white balance, (f) the plot shows signals from pixel numbers along the dotted red line in the raw image (g) original image of the 24 -patch Macbeth Color Chart.

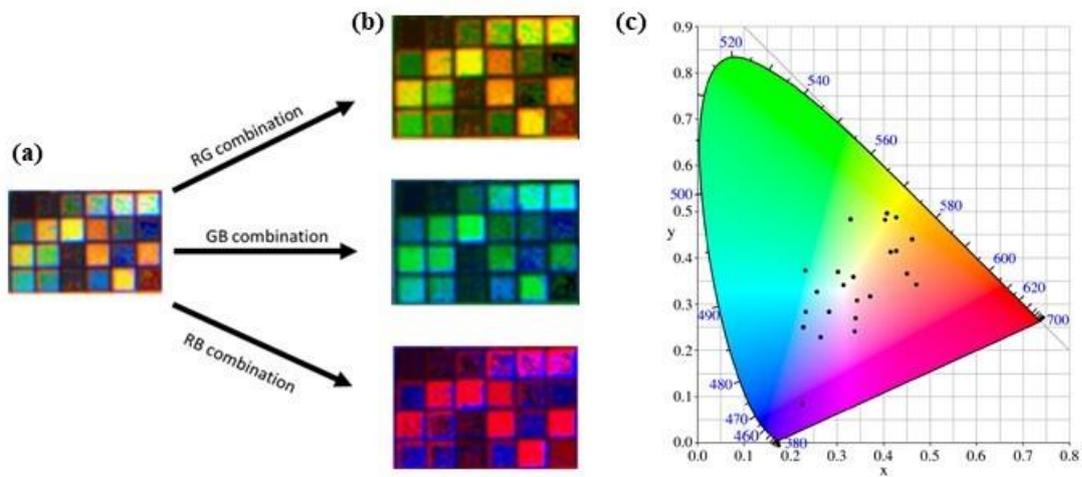

Figure 5: Demonstration of image overlay of different bands (a) recovered Macbeth Chart from the single sensor based multispectral camera from Figure 4 (b) RG, GB and RB combinations of Macbeth ColorChecker (c) CIE chart for the recovered 24-patch Macbeth colours.

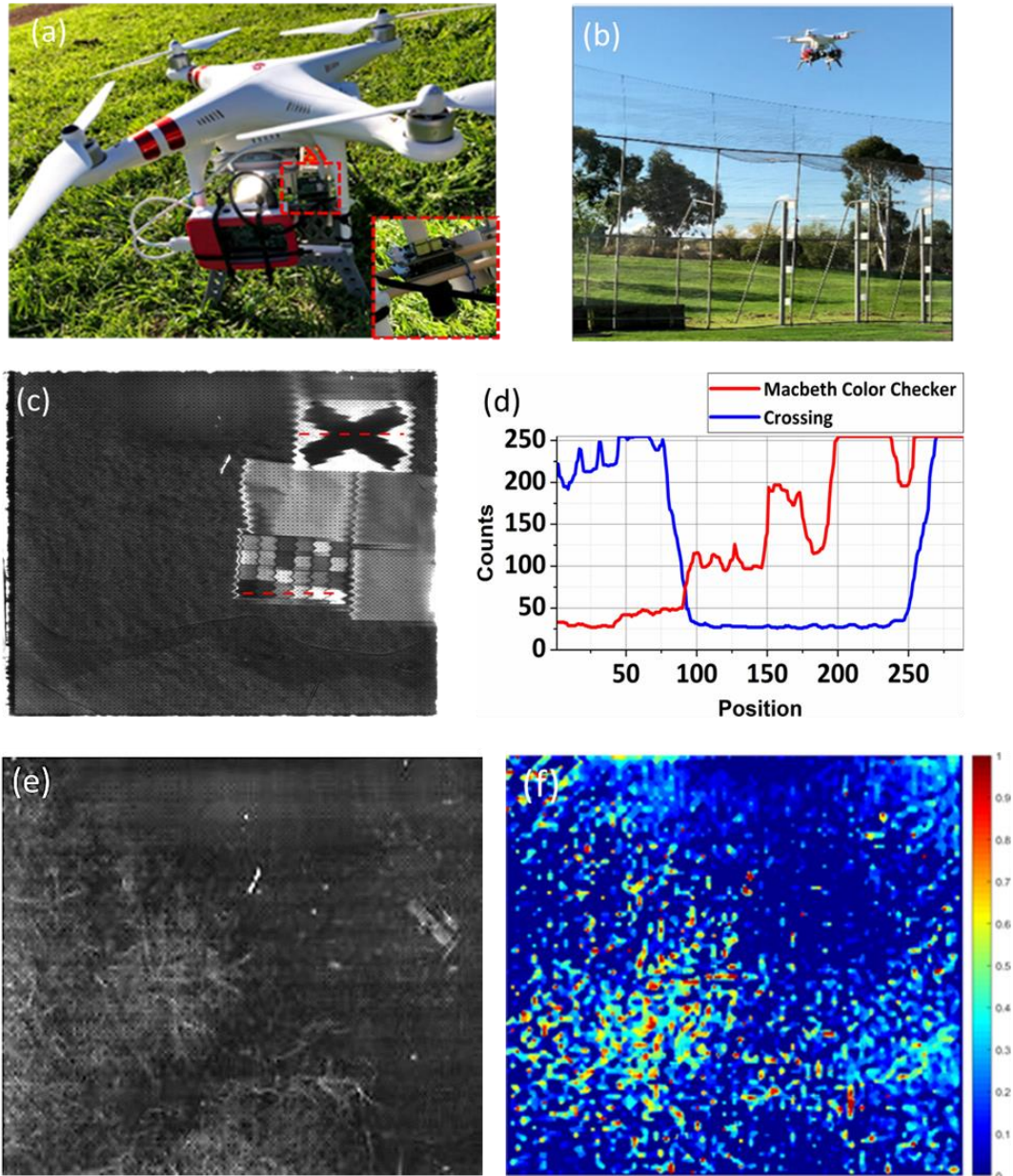

Figure 6: Demonstration of multispectral imaging of the camera using a drone platform (a) The single sensor based multispectral camera mounted on a DJI Phantom Drone without using any Gimbal (b) Image capturing using the camera (c) 8 bit raw Image captured from 15 m by the single sensor based camera showing the clarity of different patterns on ground (d) the plot shows signals from pixel numbers along the dotted red line in the raw image of crossing (blue line) and Macbeth Chart (red line) (e) 8 bit raw multispectral image of healthy grass and dry grass captured by the camera (f) R, G and B bands are recovered from the raw multispectral image to get RB vegetation index (RBVI) . RBVI images shows area of high dense green grass (red colour in the image) compared to dry grass (blue colour).